\documentclass[12pt]{article}
\usepackage{arxiv}
\usepackage[numbers]{natbib}
\usepackage{graphicx}
\usepackage{amsmath, amssymb}
\usepackage[T1]{fontenc}    
\usepackage{amsfonts}       
\usepackage{nicefrac}       
\usepackage{microtype}      

\mathchardef\mhyphen="2D

\title{Chorus Wave–Driven Electron Dynamics in the Van Allen Belts: From Coherence to Diffusion}

\author{
  Xin Tao\thanks{Correspondence to: \texttt{xtao@ustc.edu.cn}.}\ \ and Zeyu An\\
School of Earth and Space Sciences, University of Science and Technology of
China, Hefei, China,
\And
   Fulvio Zonca\\
Center for Nonlinear Plasma Science and C.R. ENEA Frascati, C.P. 65, Frascati, Italy
\And
   Liu Chen\\
Institute of Fusion Theory and Simulation and Department of
Physics, Zhejiang University, China \\
\AND
 Jacob Bortnik\\
Department of Atmospheric and Oceanic Sciences, University of California, Los Angeles, CA, USA
}

\begin{document}

\maketitle

\begin{abstract}

The Van Allen radiation belts contain relativistic electrons trapped by Earth’s
magnetic field, posing serious risks to spacecraft. Chorus waves are known to
accelerate these electrons via resonant interactions, but these interactions are
inherently nonlinear and coherent. How such processes shape large-scale electron
dynamics remains unresolved. Two competing paradigms, nonlinear advection and
diffusive transport, have been debated for decades. Here, we address this controversy
using large-scale first-principles simulations that self-consistently generate
realistic chorus wave fields, coupled with test particle modeling. We find that
electron motion is coherent on short timescales -- comparable to or less than a
bounce period -- but becomes stochastic over longer timescales due to phase
decorrelation. The resulting transport coefficients support the use of
quasilinear diffusion theory for long-term evolution. This work bridges
microscopic nonlinear physics with macroscopic modeling frameworks, offering a
unified explanation of radiation belt dynamics and advancing the foundation for
space weather forecasting.

\end{abstract}

\section{Introduction}

Earth's Van Allen radiation belts were first discovered
at the dawn of the space age in 1958 by the Explorer 1 and 3
spacecraft \cite{VanAllen1959a}. More than half a century of in-situ satellite
observations reveal that the relativistic electron flux in the dynamic outer
belt can increase by orders of magnitude during geomagnetic
storms \cite{Li1997,Reeves2003,Baker2018}. The enhanced level of radiation can
cause anomalies in spacecraft operations \cite{Baker2018}. Whistler-mode waves
are believed to be the primary drivers of acceleration of radiation belt
electrons \cite{Horne2005b,Chen2007,Thorne2013b,Reeves2013}. In the outer
radiation belt, these waves predominantly manifest as chorus
\cite{Tsurutani1974,Li2012}, consisting of repetitive, discrete, quasi-coherent
elements \cite{Santolik2003}. Global modeling of the Van Allen radiation belt
electron flux evolution often relies on quasilinear diffusion theory, which
approximates chorus waves as broadband
\cite{Horne2005b,Thorne2013b,Shprits2008b}. However, this approach is
controversial because chorus waves violate the basic assumptions of quasilinear
theory \cite{Albert2002,Bortnik2008b,Ripoll2020,Li2019a,Allanson2024}. First,
chorus waves are generated by coherent nonlinear wave-particle interactions
\cite{Omura2008,Tao2021,Zonca2022}, while quasilinear theory assumes wave
generation by a linear instability \cite{Kennel1966}. Second, the large
amplitudes and quasi-coherent nature of chorus waves lead to coherent electron
motions \cite{Albert2002,Omura2007,Bortnik2008b}, whereas quasilinear theory
assumes small-amplitude, broadband waves that result in stochastic particle
behavior and diffusive evolution of the phase space density. These coherent
motions can result in the rapid energization of radiation belt electrons to MeV
energies within seconds to minutes \cite{Albert2002,Omura2007,Artemyev2014},
which is not accounted for in diffusive models. Nonetheless, discrepancies exist
between this nonlinear acceleration timescale and observational data, as typical
variations in radiation belt electron flux during geomagnetic storms occur over
hours to days \cite{Li1997,Reeves2003}. While approximating chorus waves as
broadband may yield electron evolution timescales more consistent with
observations \cite{Horne2005b,Thorne2013b}, limited satellite coverage casts
doubt on whether this agreement merely reflects uncertainties in the wave
parameters used in radiation belt modeling \cite{Ripoll2020,Li2019a}. Addressing
these fundamentally opposing paradigms is critical for evaluating the validity
of approximating chorus waves as broadband in radiation belt modeling and for
establishing an accurate framework for predicting radiation belt dynamics, a key
element of space weather forecasting.

To address the challenges posed by the quasi-coherent nature of chorus waves,
previous efforts have incorporated nonlinear effects into radiation belt models
by treating rapid scattering and acceleration as advection or jump processes
within a transport framework \cite{Albert2002,Artemyev2014}. These nonlinear
advection terms predict much faster electron transport than quasilinear theory,
potentially leading to significant advection in phase space. However, on longer
timescales, some studies show that transport remains diffusive
 \cite{Artemyev2020}, though with diffusion rates far exceeding quasilinear
estimates, reducing the characteristic timescale to minutes \cite{Mourenas2018}.
These conclusions often rely on simplified synthetic models of chorus waves that
overlook key realistic properties. This limitation arises because satellite
observations, constrained to single-point measurements, cannot capture the full
spatial and temporal evolution of chorus waves. Moreover, when repetitive chorus
elements are considered, results from test particle simulations are also
debated, as synthetic wave fields frequently treat these elements as overly
uniform \cite{Saito2012, Tao2014d}. Consequently, the long-standing debate over
whether chorus-driven electron transport is dominantly advective or diffusive,
and what timescales govern the resulting electron dynamics, remains unresolved.

In this work, we employ self-consistent particle-in-cell (PIC) simulations based
on first principles to generate chorus wave fields with realistic
features \cite{Tao2014b,Tao2017b}, overcoming the limitations of synthetic wave
models used in previous test particle studies. By reproducing key observational
properties of chorus waves and coupling the resulting wave fields with test
particle simulations, we demonstrate that the nature of electron–chorus wave
interactions depends on the timescale of interest. While interactions are
coherent on short timescales comparable to the duration of a single chorus
element, electron dynamics on timescales longer than a few bounce periods become
equivalent to those driven by a broadband wave field.

\section{Results}

\subsection{PIC simulation of chorus waves}

\begin{figure}[htb]
  \centering
  \includegraphics[width=1.0\textwidth]{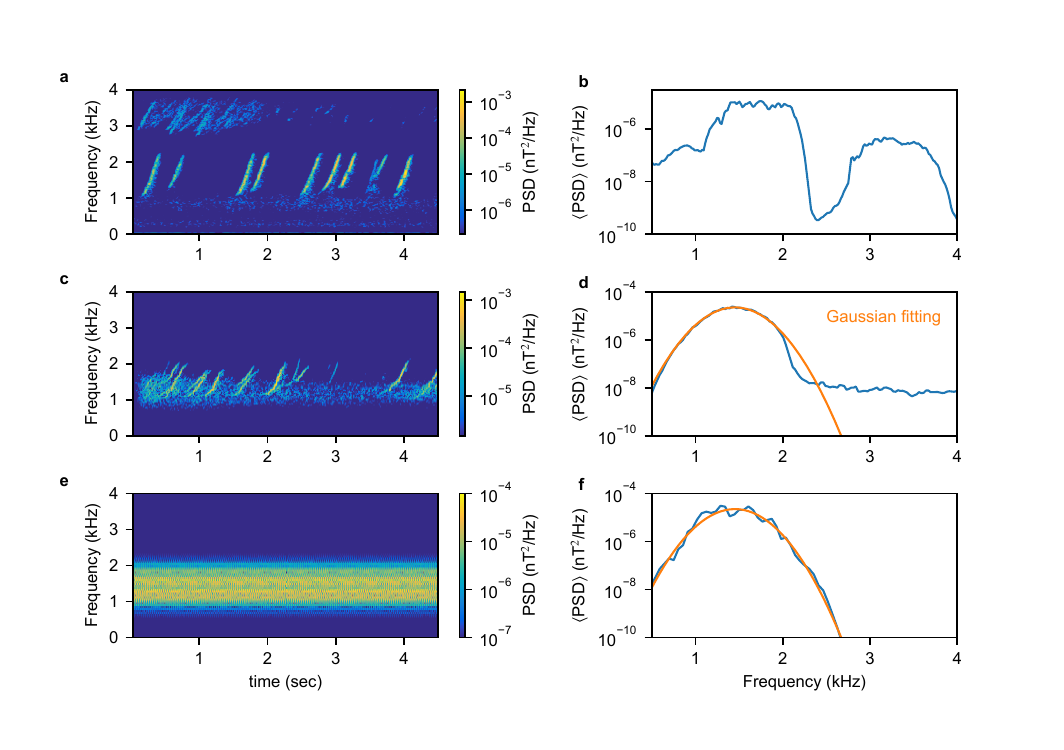}
  \caption{\label{fig_wave_spt} \textbf{Typical wave fields from Van Allen
  Probes \cite{Mauk2013a} observations and computer simulations.} \textbf{a},
Whistler-mode chorus waves observed by Van Allen Probe A at $L = 5.2$ and
magnetic latitude $\text{MLAT}=-1.2^{\circ}$ on 2014-04-30 UT, with the
corresponding time-averaged power spectral density, $\langle \text{PSD}\rangle$,
displayed in \textbf{b}. \textbf{c}, The chorus wave field from a computer
simulation at an equivalent latitude of $\lambda = 3^\circ$, with the
time-averaged power spectral density in \textbf{d} fitted by a Gaussian (orange
line). \textbf{e}, The constructed broadband wave field, whose power spectral
density (\textbf{f}) is derived from the Gaussian fit, as indicated. }
\end{figure}

Previous studies have shown that PIC simulations can effectively replicate the
observed characteristics of individual chorus elements \cite{Zhang2021},
including narrowband quasi-coherent structures, dynamic amplitude
evolution \cite{Tao2012a}, and phase decoherence \cite{Zhang2020}. Using the
latest simulation techniques \cite{Lu2021}, we generate repetitive chorus
elements with a typical repetition period of approximately 0.4 seconds,
consistent with observations in the magnetosphere \cite{Shue2015}. The
simulations use a background magnetic field corresponding to an equatorial
radial distance of $5$ Earth radii in Earth's dipole field, which is a typical
outer radiation belt region, with parameters chosen to represent lower-band
chorus waves. This setup ensures that the generated wave fields closely resemble
those observed in situ. As shown in Figure \ref{fig_wave_spt}b, the simulated
spectrograms successfully capture the key narrowband, repetitive, and dynamic
characteristics of chorus. A comparison with observational data in Figure
\ref{fig_wave_spt}a reveals similar, though not identical, elements and chirping
rates. Due to computational constraints, we assume parallel propagation and
neglect field line curvature but still retain field line inhomogeneity. However,
neglecting oblique propagation does not affect our main conclusions, because
observed chorus waves are predominantly quasi-parallel. Furthermore, oblique
waves experience Landau damping as they propagate to higher
latitudes \cite{Bortnik2006} and require larger amplitudes to initiate nonlinear
interactions \cite{Tao2010d}. Most importantly, if such nonlinear interactions
occur, the decorrelation process described in this study applies equally well.
As such, our study represents a ``worst-case scenario'', where coherent,
parallel-propagating waves are the most difficult to decorrelate. To assess the
validity of previous global radiation belt models, we construct a broadband wave
field for comparison by applying a Gaussian fit to the time-averaged power
spectral density of the simulated chorus waves (Figure \ref{fig_wave_spt}d-f), a
common approach in transport coefficient calculations \cite{Meredith2012}. Test
particle simulations using both the chorus and broadband wave fields allow us to
evaluate and compare their respective effects.

\subsection{Test particle simulations}

\begin{figure}
  \centering 
  \includegraphics{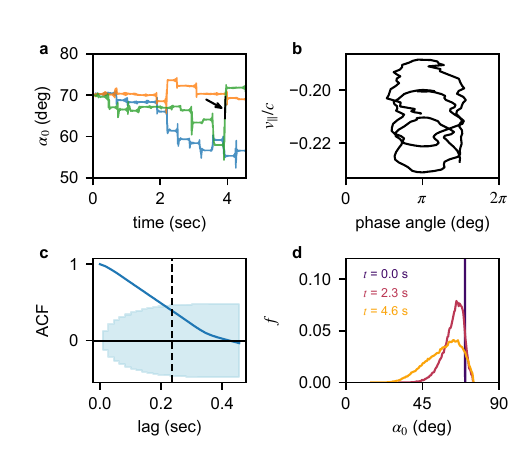}
  \caption{\label{stochasticity}\textbf{Time-scale dependent behavior of
    electrons interacting with chorus.} \textbf{a}, Three sample trajectories of
    electrons with initial energy $E/E_{0} = 0.2$ and initial equatorial pitch
    angle $\alpha_{0} = 70^{\circ}$. The detailed parameters used in the test
    particle simulations are provided in the Appendix. \textbf{b}, Phase
    space trajectory of a phase-trapped electron, with its orbit highlighted in
    black and indicated by an arrow in \textbf{a}. \textbf{c},  Autocorrelation
    function of the $\alpha_{0}$ time series for the trajectory shown
    in green in \textbf{a}. The shaded region marks the 95\% confidence interval
    under the null hypothesis of white noise, indicating the range within which
    autocorrelation values are not statistically significant. The vertical black dashed
    line indicates the time lag at which the autocorrelation function decays to
    $1/e$. \textbf{d}, Evolution of the electron pitch angle distribution
    function $f(\alpha_0)$ at three times ($t = 0.0$\,s, 2.3\,s, and
    4.6\,s), showing progressive broadening consistent with diffusive transport.
  }
\end{figure}

We begin by investigating whether the long-term dynamics of electrons
interacting with chorus waves exhibit diffusive behavior. To this end, we
perform test particle simulations using chorus wave fields generated
self-consistently from PIC simulations. Figure~\ref{stochasticity} presents
representative electron trajectories and the temporal evolution of their
distributions for electrons with initial equatorial pitch angle $\alpha_{0} =
70^{\circ}$ and energy $E/E_{0} = 0.2$, where $E_{0} = 0.511$\,MeV is the
electron rest energy. In Figure~\ref{stochasticity}a, the equatorial pitch angle
exhibits noticeable coherent variations over portions of the displayed orbits.
One such trajectory, replotted in phase space, displays a characteristic
circling pattern indicative of nonlinear phase trapping
(Figure~\ref{stochasticity}b). However, due to the discrete nature of chorus
elements, the timescale of such trapping is limited. The outcomes of
interactions with two consecutive chorus elements are effectively uncorrelated,
as the effects of nonlinear wave–particle interactions depend sensitively on the
initial interaction phase angle. In general, electrons may gain pitch angle and energy
if phase-trapped or lose them if untrapped. Additionally, realistic features of
chorus waves—such as amplitude modulation  \cite{Tao2012a,Tao2013} and phase decoherence
 \cite{Zhang2020}—captured in the PIC-generated fields enhance the stochastic
nature of the dynamics and diminish the efficiency of nonlinear advective
acceleration \cite{An2022}. These stochastic features are also evident in the trajectories
shown in Figure~\ref{stochasticity}a.

To quantify the duration of coherence, we compute the autocorrelation function
of the pitch angle time series for the trajectory shown in green in
Figure~\ref{stochasticity}a.  As shown in Figure~\ref{stochasticity}c, the
autocorrelation falls below the 95\% confidence bounds (shaded region) at a lag
of approximately 0.2 seconds and remains within the bounds thereafter,
indicating that pitch angle variations become statistically uncorrelated beyond
this timescale. These confidence bounds represent the expected range of
autocorrelation values under the null hypothesis of white noise. The
decorrelation timescale—defined as the lag where the autocorrelation decays to
$1/e$—is approximately 0.24 seconds, or about $0.4 \tau_{b}$, where $\tau_{b} =
0.61$\,seconds is the bounce period for the selected energy and pitch angle.
While this timescale varies among electrons, interactions are generally coherent
up to approximately $\mathcal{O}(\tau_{b}/4)$, since chorus waves are generated
near the equator and propagate toward higher latitudes, and resonant electrons
must travel opposite to the wave propagation direction. Therefore, when
electron trajectories are sampled at intervals significantly longer than
$\tau_{b}/4$, their dynamics appear approximately Markovian and stochastic.
Although the process is more complex, it is fundamentally similar to classical
Brownian motion: a particle's motion is deterministic over short timescales
(between collisions), but becomes effectively stochastic over timescales much
longer than the decorrelation time \cite{Einstein1905Brownian}. Figure~\ref{stochasticity}d shows the
evolution of the pitch angle distribution function $f$ at $t = 0.0$\,s, 2.3\,s,
and 4.6\,s. The progressive broadening of the distribution over time is
indicative of diffusive rather than convective evolution. More detailed
comparisons with broadband wave simulations and predictions from quasilinear
theory will be presented in later sections.

\subsection{Comparison of transport coefficients}

\begin{figure}
\centering
\includegraphics[width=1.0\textwidth]{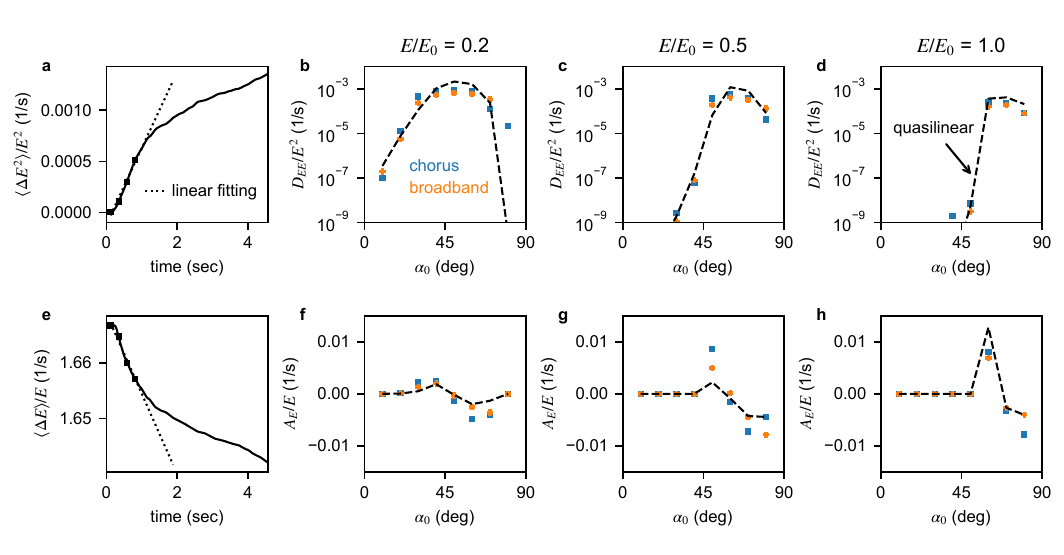}
\caption{\label{cmp_trans_coeff}\textbf{Comparison of transport coefficients
from chorus waves, broadband waves, and quasilinear theory.} \textbf{a}, A
sample showing variation of the variance of the energy of test particles, whose
initial $\alpha_{0} = 70^{\circ}$ and $E/E_{0} = 0.5$. The test particle
diffusion coefficient $D_{EE}$ is half of the slope of the linear fitting
(dotted line) to the time rate change of the variance of the energy distribution
(solid line). \textbf{b}, Comparison of diffusion coefficients $D_{EE}/E^{2}$
from test particle calculations using the chorus wave field (blue squares),
broadband wave field (orange crosses), and quasilinear theory (black dashed
lines) for the initial energy $E/E_{0} = 0.2$ at eight equidistant pitch angles
ranging from $10^{\circ}$ to $80^{\circ}$. \textbf{c} and \textbf{d} are similar
to \textbf{b} but for initial energies $E/E_{0} = 0.5$ and $E/E_{0} = 1.0$,
respectively. \textbf{e}, A sample showing the variation of the mean of the
energy, $\langle E \rangle$, of all test particles. The test particle advection
coefficient is the slope of the linear fitting to the time rate change of
$\langle E \rangle$. \textbf{f}–\textbf{h}, Similar to \textbf{b}–\textbf{d},
but comparing advection coefficients $A_{E}/E$.}

\end{figure}

The system's stochastic nature allows us to quantify the long-term transport of
radiation belt electrons using Fokker-Planck equations, with diffusion rates and
advection rates serving as key parameters. We calculate the diffusion rates,
$D_{EE}$, and advection coefficients, $A_{E}$, for energy ($E$), as shown in
Figure~\ref{cmp_trans_coeff}. These energy transport coefficients enable a
direct estimation of the electron energization timescale. Other transport
coefficients, such as those related to equatorial pitch angle, are directly
linked to $D_{EE}$ and $A_{E}$ and are therefore not shown. We compare results
for three different energies $E/E_{0} = 0.2, E/E_{0}= 0.5$, and $E/E_{0}=1.0$.
Diffusion and advection coefficients are determined by linearly fitting the
variance and mean of $E$ across all test particles in the simulations. To
minimize statistical noise, we use 10,000 test particles for each simulation,
recording the mean and variance every $ \tau_{b}/2 $, starting from $ t =
\tau_b/4 $. Sampling at intervals of $ \tau_b/2 $ ensures decorrelation of
successive data points.

We compare transport coefficients for chorus waves with those for broadband
waves and quasilinear theory, as shown in Figure~\ref{cmp_trans_coeff}. The
transport coefficients for chorus and broadband waves exhibit strong agreement,
except near $\alpha_{0} = 90^{\circ}$ for $E/E_{0} = 0.2$, where differences in
the wave power spectral density (Figures~\ref{fig_wave_spt}d and
\ref{fig_wave_spt}f) account for the discrepancy. This agreement demonstrates
that the long-term dynamics of electrons interacting with quasi-coherent chorus
waves can be effectively modeled using an equivalent broadband wave field.
Additionally, the comparison between quasilinear transport coefficients and
those from test particle simulations shows good overall consistency, despite
some differences. For instance, the maximum quasilinear diffusion coefficients
exceed test particle results by a factor of 2 to 3, while for $E/E_{0} = 0.5$
the maximum quasilinear advection coefficients are smaller by a factor of 2 to 4
at $\alpha_{0}=50^{\circ}$. These differences likely result from the simplifying
assumptions of quasilinear theory, as discussed later in the context of
distribution function comparisons. Overall, the transport coefficients for
chorus waves, broadband waves, and quasilinear theory remain largely consistent.
Consequently, our findings do not support significant secular long-term
nonlinear advection arising from quasi-coherent nonlinear interactions between
electrons and chorus waves, despite the existence of such advection on
timescales comparable to or shorter than $\tau_{b}/4$.

\subsection{Comparison of distribution functions}
\begin{figure}
\centering
\includegraphics[width=1.0\textwidth]{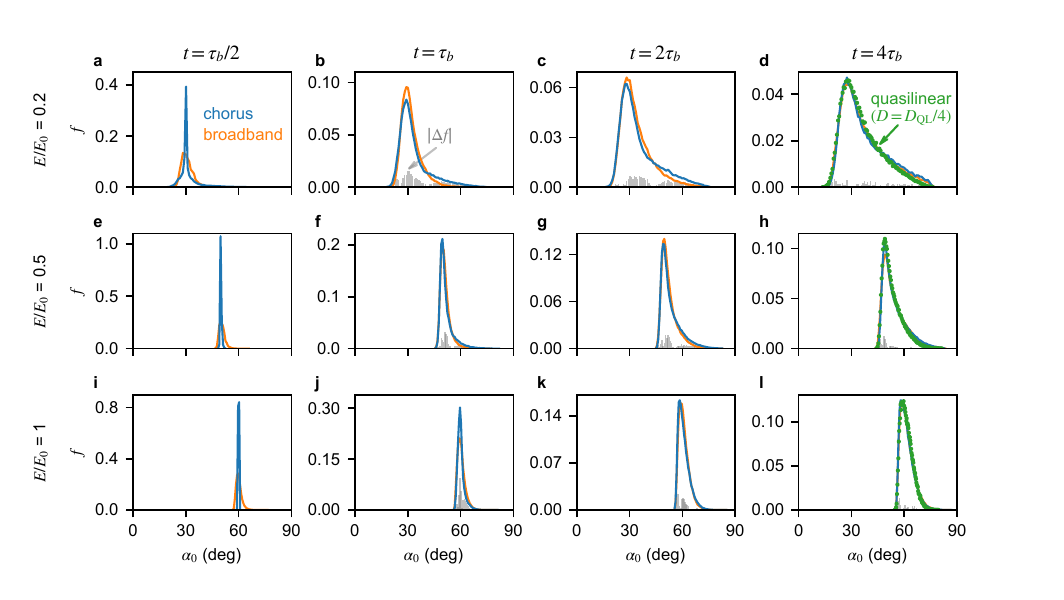}
  \caption{\label{cmp_dist}\textbf{Comparison of electron equatorial pitch angle
  distributions.} \textbf{a}, Electron pitch angle distribution at $t =
\tau_{b}/2$ for an initial energy of $E/E_{0} = 0.2$ and an initial pitch angle
$\alpha_{0} = 30^{\circ}$. Test particle distributions from interactions with
chorus waves (blue) are compared with those from broadband waves (orange).
\textbf{b}–\textbf{d}, Similar to \textbf{a}, but for $t = \tau_{b}$,
$2\tau_{b}$, and $4\tau_{b}$, respectively. Grey bars indicate the absolute
difference between the distributions from chorus and broadband waves, $|\Delta
f|$. Green dots represent distributions predicted by quasilinear theory using $D
= D_\text{QL}/4$. \textbf{e}–\textbf{h}, Similar to \textbf{a}–\textbf{d}, but
for an initial energy of $E/E_{0} = 0.5$ and an initial pitch angle $\alpha_{0}
= 50^{\circ}$. \textbf{i}–\textbf{l}, Similar to \textbf{a}–\textbf{d}, but for
an initial energy of $E/E_{0} = 1$ and an initial pitch angle $\alpha_{0} =
60^{\circ}$.}

\end{figure}

To make further comparisons between the chorus and broadband waves, we show the
evolution of the distribution functions from test particle simulations for the
three different energies and pitch angles in Figure~\ref{cmp_dist} at
$t=\tau_{b}/2, \tau_{b}, 2\tau_{b}$ and $4\tau_{b}$. At $t = \tau_{b}/2$, the
distributions from chorus and broadband waves differ significantly due to the
discrete nature of the chorus elements. A substantial fraction of electrons
remains unscattered by chorus, resulting in a peak near the initial pitch angle.
In contrast, broadband waves continuously scatter electrons from the beginning.
The discrete nature of chorus also contributes to the differences in the
distribution near the initial $\alpha_{0}$ at other times. By $t = \tau_{b}$ and
$t = 2\tau_{b}$, the distributions from both wave types converge, although the
distribution from chorus waves is slightly larger than that from broadband waves
at pitch angles $\alpha_{0}$ near $60^{\circ}$ for $E/E_{0} = 0.2$ and $E/E_{0}
= 0.5$. These differences are the effects of coherent nonlinear scattering,
which can cause a large increase in electrons' pitch angle from a single
interaction. At $t = 4\tau_{b}$, the chorus and broadband wave distributions
become nearly indistinguishable. These results demonstrate again that the
long-term dynamics of electrons from interactions with chorus can be well
approximated by a broadband wave field due to decorrelation between successive
interactions.

Figure~\ref{cmp_dist} also presents comparisons of the electron distribution
with predictions from quasilinear theory at $t = 4\tau_{b}$. We solve the
Fokker-Planck equation using a Monte Carlo method based on stochastic
differential equations \cite{Tao2008}. Using the theoretical diffusion
coefficients ($D_{\text{QL}}$) results in stronger diffusion than observed in
test particle simulations, as expected from comparisons of transport
coefficients. Adjusting the diffusion coefficients to $D_{\text{QL}}/2$ and
$D_{\text{QL}}/4$ shows that $D_{\text{QL}}/4$ achieves excellent agreement with
the test particle results; therefore, only theoretical distributions using
$D_{\text{QL}}/4$ are shown in the figure. Comparisons at other times are not
shown for clarity of presentation, but similar agreement exists between the
quasilinear and broadband wave results. It is important to note that the
requirement for reduced diffusion coefficients is not due to coherent nonlinear
interactions but rather to the large amplitude of the broadband waves.
Quasilinear theory is designed for small amplitude waves. When the amplitude of
broadband waves exceeds a certain threshold, the actual diffusion coefficients
may become saturated and fall below the predictions of quasilinear theory. The
exact threshold amplitude depends on the wave and particle parameters. For
typical radiation belt conditions, a rough estimate from a previous
study \cite{Tao2012c} suggests that the normalized average wave amplitude,
$\delta B/B_{0}$ (where $B_{0}$ represents the background magnetic field),
should be less than $10^{-3}$ for quasilinear theory to be accurate. In our
simulations, the normalized average amplitude of both broadband and chorus waves
exceeds $10^{-3}$  at latitudes greater than $5^{\circ}$. Correspondingly,
diffusion coefficients from quasilinear theory are expected to exceed those from
test particle calculations, as shown in Figure~\ref{cmp_trans_coeff}. A set of
reduced theoretical diffusion coefficients is therefore needed to model the
evolution of the distribution function, as illustrated in Figure~\ref{cmp_dist}.
If the equivalent broadband waves had a smaller amplitude, quasilinear theory
would more closely match test particle simulations for both wave types.

\section{Discussion and Summary}
Together, these calculations reveal the complex dynamics of electron behavior
arising from interactions with chorus waves. For timescales on the order of or
shorter than $\tau_{b}/4$, nonlinear coherent scattering by chorus significantly
influences electron distributions, especially when chorus wave amplitudes are
large. Such processes include chorus wave excitation and electron precipitation
through scattering by individual chorus elements. However, over timescales much
longer than the bounce period, the loss of coherence between successive
scattering allows electron motion to be treated as stochastic. Our results
indicate that, for typical chorus wave parameters in the Van Allen belts, the
diffusion coefficients and long-term evolution of the electron distribution
closely match those of an equivalent broadband wave field. Notably, our results
do not support the presence of rapid nonlinear advection or of significantly
enhanced diffusion on extended timescales.

It is important to note that, strictly speaking, this diffusion is not
quasilinear, as chorus wave excitation involves a coherent nonlinear process.
However, the quasilinear diffusion theory commonly applied in radiation belt
modeling is in the resonant limit, where Van Allen belt electrons are considered
to be test particles. This assumption is justified because chorus waves are
mainly generated by a denser population of lower-energy keV electrons, while the
relatively low density of Van Allen belt electrons (a few hundred keV to MeV)
minimizes their feedback on chorus wave growth and are thus considered to be a
parasitic population. The strong agreement between results for chorus and
broadband waves, therefore, supports the use of quasilinear theory as a
foundational framework for radiation belt modeling in space weather prediction,
provided that equivalent broadband wave amplitudes remain sufficiently small for
quasilinear theory to yield accurate predictions.

\appendix

\section{Computer Simulation of Chorus Waves}
We use a previously developed computer code, DAWN \cite{Tao2014b,Tao2017b},
which solves the Vlasov equation coupled with Maxwell’s equations, to
self-consistently simulate the generation of chorus waves in Earth's
magnetosphere. To simulate the repetitive generation of chorus elements, we
solve the following modified Vlasov equation, based on a recent study
\cite{Lu2021}:
\begin{align}
  \frac{\partial f}{\partial t} + \dot{z}\cdot \frac{\partial f}{\partial z} +
  \dot{\boldsymbol{u}}\cdot \frac{\partial f}{\partial \boldsymbol{u}} =
  -\frac{f-f_{0}}{\tau}.
\end{align}
In this equation, $f$ is the phase space density, $z$ is the spatial coordinate,
and $\boldsymbol{u} = \gamma \boldsymbol{v}$, where $\boldsymbol{v}$ is the
particle velocity and $\gamma$ is the relativistic factor. The time derivative
of $z$ and $\boldsymbol{u}$ are denoted by $\dot{z}$ and $\dot{\boldsymbol{u}}$,
respectively, with $\dot{\boldsymbol{u}}$ given by the Lorentz force equation.
The term $-(f-f_{0})/\tau$ on the right-hand side simulates the continuous
injection of fresh energy into the system by restoring the total distribution to
the initial state, providing free energy to generate new chorus elements.  The
characteristic repetition period $\tau$ is set to $25000\,\Omega_{e0}^{-1}$ in
normalized units, where $\Omega_{e0}$ is the equatorial electron cyclotron
frequency, equivalent to approximately 0.57 seconds in SI units. This setting
produces chorus elements with an average repetition period of about 0.4 seconds,
consistent with typical observations in the magnetosphere \cite{Shue2015}.

In the DAWN code, the electron distribution includes two components: a cold
component modeled by fluid equations and a hot component with a bi-Maxwellian
distribution. The hot component has temperatures of $10.12$\,keV and $14.1$\,keV
along and perpendicular to the magnetic field, respectively. These temperature
values are chosen specifically to generate chorus elements in the lower band.
The cold electron number density is chosen to satisfy $\omega_{pe}/\Omega_{e0} =
5$, where $\omega_{pe}$ is the electron plasma frequency, representing typical
conditions outside the plasmapause. The hot electron density is set at $0.6\%$
of the cold electron density. Ions are considered fixed, as the frequency of
chorus waves is much higher than the ion cyclotron frequency.

The dominant component of the background magnetic field is represented by $B_{z}
= B_{z0} (1+\xi z^{2})$, an approximation of Earth’s dipole field without
considering field line curvature. The $B_{x}$ and $B_{y}$ components are given
by $B_{x} = -(x/2)\mathrm{d} B_{z}/\mathrm{d} z$ and $B_{y} = -(y/2)\mathrm{d}
B_{z}/\mathrm{d} z$ to ensure $\nabla\cdot \boldsymbol{B} = 0$. Here, $\xi$
characterizes magnetic field inhomogeneity. For a dipole field, $\xi = 4.5/(L
R_{E})^{2}$, with $L$ being the $L$-shell and $R_{E}$ the Earth radius
\cite{Helliwell1967}. We set $\xi = 4.43\times10^{-15}$\,m$^{-2}$ to represent
Earth's dipole field at $L=5$, which corresponds to the central region of the
outer radiation belt. Our simulation domain covers both hemispheres, with the
maximum latitude being about $15^{\circ}$, consistent with typical nightside
latitudinal distributions of chorus waves from observation.

\section{Test Particle Simulations}

Two types of wave fields are used in the test particle simulations: the
self-consistent PIC-generated chorus waves and a constructed broadband wave
field. The PIC-generated chorus waves are saved with a time step of $0.4
\Omega_{e0}^{-1}$. For lower-band chorus waves with $\omega/\Omega_{e0} = 0.3$,
this time step provides 50 data points within one wave period. The broadband
wave field is constructed by superposing 100 single-frequency waves
\cite{Tao2012c}, with frequencies evenly distributed between $0$ and
$0.5\,\Omega_{e0}$. These broadband waves propagate from the equator to both
hemispheres. The latitude-dependent power spectral density of the broadband
waves is determined from the chorus wave field at 25 latitudes, equally spaced
between the equator and the maximum latitude.

In test particle simulations, we use the same background magnetic field model as
in the PIC simulation. The Boris method is used to solve the Lorentz equations
of motion with a time step of $0.02 \Omega_{e0}^{-1}$, ensuring accurate
resolution of the electron cyclotron motion. For each simulation with a given
initial energy and equatorial pitch angle, we use $10,000$ test particles to
reduce statistical noise. These test particles are randomly distributed in
initial cyclotron and bounce phase. The diffusion coefficient $D_{EE}$ and the
advection coefficient $A_{E}$ of the energy $E$ are defined as
\begin{align}
  D_{EE} = \frac{\langle \Delta E^{2}\rangle}{2 t}, \quad  A_{E} = \frac{\langle
  \Delta E\rangle}{t}.
\end{align}
where $\langle \Delta E^{2}\rangle$ and $\langle \Delta E\rangle$ are the
variance and mean of the $E$ distribution at time $t$. By linearly fitting the
variance and mean as functions of $t$, we obtain $D_{EE}$ and $A_{E}$ as
\begin{align}
  D_{EE} = k_{D}/2; \quad A_{E} = k_{b},
\end{align}
where $k_{D}$ and $k_{b}$ are the slopes of the variance and mean fits,
respectively.

\section{Solving the quasilinear diffusion equation}

The quasilinear diffusion equation used here is a 2D bounce-averaged diffusion
equation in equatorial pitch angle $\alpha_{0}$ and momentum $p$,
\begin{align}
  \frac{\partial f}{\partial t} = \frac{1}{G} \frac{\partial }{\partial \alpha_{0}} G \left(D_{\alpha_{0}\alpha_{0}} \frac{\partial f}{\partial \alpha_{0}} + D_{\alpha_{0} p} \frac{\partial f}{\partial p}\right) + \frac{1}{G} \frac{\partial }{\partial p} G \left( D_{\alpha_{0} p}\frac{\partial f}{\partial \alpha_{0}} + D_{pp} \frac{\partial f}{\partial p}\right).
\end{align}
This equation is equivalent to the following stochastic differential equations
(SDEs),
\begin{align}
  \mathrm{d} \alpha_{0} &= A_{\alpha_{0}} \mathrm{d} t + \sigma_{11} \mathrm{d} W_{1} + \sigma_{12} \mathrm{d} W_{2}, \\
  \mathrm{d} p &= A_{p} \mathrm{d} t + \sigma_{21} \mathrm{d} W_{1} +
  \sigma_{22} \mathrm{d} W_{2},
\end{align}
where the advection coefficients, $A_{\alpha_{0}}$ and $A_{p}$, are given by
\begin{align}
  A_{\alpha_{0}} &= \frac{1}{G} \frac{\partial }{\partial \alpha_{0}} \left( G
  D_{\alpha_{0}\alpha_{0}}\right) + \frac{1}{G} \frac{\partial }{\partial p}
    \left( G D_{\alpha_{0}p}\right), \\ A_{p} &= \frac{1}{G} \frac{\partial
    }{\partial \alpha_{0}} \left( G
    D_{\alpha_{0} p}\right) + \frac{1}{G} \frac{\partial }{\partial p}\left( G
  D_{pp}\right).
\end{align}
We choose $\sigma_{21}=0$ and define the other components as \cite{Tao2008}
\begin{align}
  \sigma_{11} &= \sqrt{2 D_{\alpha_{0}\alpha_{0}}},  \\
  \sigma_{21} &= \sqrt{2 D_{\alpha_{0}p}^{2} / D_{\alpha_{0}\alpha_{0}}}, \\
  \sigma_{22} &= \sqrt{2 D_{pp} - \sigma_{21}^{2}}.
\end{align}
Both $\mathrm{d} W_{1}$ and $\mathrm{d} W_{2}$ are increments of standard
Brownian motion, calculated numerically as
\begin{align}
  \mathrm{d} W(t) = \sqrt{\mathrm{d} t} N(0,1).
\end{align}
Here $N(0,1)$ is a standard Gaussian random number with zero mean and unit
variance. To solve the quasilinear diffusion equation for a given initial
$\alpha_{0}$ and $E$, we sample stochastic trajectories of 10,000 particles and
obtain the distribution function $f$.

\section*{Open Research}

The amount of data generated by the high-resolution PIC simulations and test
particle simulations is approximately 1.3 TB. Interested readers can contact
the corresponding authors to make arrangements for the transfer of those data.

\bibliographystyle{plainnat}  
\bibliography{refs}

\section*{Acknowledgements}
This work was supported by National Science Foundation of China grants (42474218 and 42174182).

\end{document}